# The open LPC Paul trap for precision measurements in beta decay

P. Delahaye[1,2], G. Ban[2], M. Benali[2], D. Durand[2], X. Fabian[3], X. Fléchard[2], M. Herbane[2], E. Liénard[2], F. Mauger[2], A. Méry[4], Y. Merrer[2], O. Naviliat-Cuncic[2,5], G. Quéméner[2], B. M. Retailleau[1], D. Rodriguez[6], J. C. Thomas[1], P. Ujic[1]

[1]GANIL, CEA/DSM-CNRS/IN2P3, Bd Henri Becquerel, 14000 Caen, France

[2]Normandie Univ, ENSICAEN, UNICAEN, CNRS/IN2P3, LPC Caen, 14000 Caen, France

[3]Institut de Physique nucléaire de Lyon, 4 rue Enrico Fermi, 69622 Villeurbanne, France

[4]CIMAP, CEA/CNRS/ENSICAEN, Université de Caen, Caen, France

[5]National Superconducting Cyclotron Laboratory and Department of Physics and Astronomy, Michigan State University, East Lansing MI 48824, USA

[6]Departamento de Física Atómica Molecular y Nuclear, Universidad de Granada, Granada, Spain

Corresponding author: delahaye@ganil.fr

**Abstract.** The LPCTrap experiment uses an open Paul trap which was built to enable precision measurements in the beta decay of radioactive ions. The initial goal was the precise measurement of the beta-neutrino angular correlation coefficient in the decay of $^6$He. Its geometry results from a careful optimization of the harmonic potential created by cylindrical electrodes. It supersedes previously considered geometries that presented a smaller detection solid angle to the beta particle and the recoiling ion. We describe here the methods which were used for the potential optimization, and we present the measured performances in terms of trapping time, cloud size and temperature, and space charge related limits. The properties of the ion cloud at equilibrium are investigated by a simple numerical simulation using hard sphere collisions, which additionally gives insights on the trapping loss mechanism. The interpretation for the observed trapping lifetimes is further corroborated by a model recently developed for ion clouds in Paul traps. The open trap shall serve other projects. It is currently used for commissioning purpose in the TRAPSENSOR experiment and is also considered in tests of the Standard Model involving the beta decay of polarized $^{23}$Mg and $^{39}$Ca ion in the frame of the MORA experiment. The latter tests require in-trap polarization of the ions and further optimization of the trapping and detection setup. Based on the results of the simulations and of their interpretation, different improvements of the trapping setup are discussed.

# 1. Introduction

The LPCTrap experiment measures the β-ν angular correlation in the β decay of radioactive ions for tests of the Standard Model [1]. It uses a transparent Paul trap which allows the detection of the recoiling ion and beta particle in coincidence. As a rather natural extension of LPCTrap experiments carried out so far, the so-called *Matter's Origin from the RadioActivity of trapped and oriented ions* (MORA) project aims at measuring the CP violating *D*-triple correlation (see for example [2]) of laser - polarized $^{23}$Mg and $^{39}$Ca ions confined in an open Paul trap. Using such a laser-polarization technique in a LPCTrap-like setup would additionally enable precise measurements of other correlations involving oriented nuclei such as the beta asymmetry $A_\beta$ or the neutrino asymmetry $B_\nu$ [2]. MORA will require the highest statistics to search for New Physics (NP) with the highest sensitivity. In this regard, the current open trap presents some limitations. In particular, the trapping lifetime is limited to a fraction of second, which causes unwanted losses and background events for the isotopes best suited to this measurement. After recalling the simple procedure which was followed to optimize the trapping of ions in the current open trap, the explanation for the short trapping times observed so far is being numerically investigated. Different methods for increasing the trapping time are then proposed for the *D*-correlation measurement.

## 2. The original open trap

In 3D Paul traps, ions are confined by means of a quadrupolar RadioFrequency (RF) potential [3]. In an "ideal" Paul trap, i.e. generating a pure quadrupolar potential, the trapping potential can be expressed using the $(r,z)$ cylindrical coordinates in the following way:

(1) $\quad V_{ideal}(r,z,t) = V_{rf}(t) \cdot \frac{r^2 - 2z^2}{2r_0^2}$

where $r_0$ is a distance parameter, and $V_{rf}(t)$ is a sum of a Direct Current (DC) potential and Alternating Current (AC) potential. The DC component is usually used for a mass selective confinement, and reduces the trapping efficiency. For a non-mass-selective confinement, a pure AC potential is therefore preferred:

(2) $\quad V_{rf}(t) = V_0 \cdot \cos(\Omega t).$

where $\Omega = 2\pi \nu_{rf}$ is the angular radiofrequency. The potential in Eq. (1) is quite commonly approximated by electrodes whose surfaces form truncated hyperboloids, consisting of one ring and two end caps. These surfaces are represented in Fig. 1. In this configuration, the distance $r_0$ corresponds to the shortest distance from the trap center to the ring electrode, and $z_0 = \frac{r_0}{\sqrt{2}}$ corresponds to the distance to the end caps. Other geometries can approximate a pure quadrupolar potential. In full generality, the potential generated by a Paul trap consisting of electrodes of finite size and/or of approximated shapes can be expressed as an infinite sum of harmonics. In spherical coordinates $(\rho, \theta, \varphi)$ this gives:

(3) $\quad V(\rho, \theta, \varphi, t) = V_{rf}(t) \cdot \sum_{n=0}^{\infty} \sum_{m=-n}^{n} C_{nm} \left(\frac{\rho}{r_0}\right)^n P_n^m(\cos\theta) e^{im\varphi}$

where the complex harmonic coefficient $C_{nm}$ represents the respective strengths of the multipoles $(n, m)$, and $P_n^m(\cos\theta)$ is the associated Legendre polynomial of order $m$ and degree $n$. The effect of the multipoles of order higher than 2 is to create instabilities in the ion motion, leading the ions out of the trap (see for example [4]). Different geometries of Paul traps were studied and compared to optimize the trapping performances for the measurement of the β-ν angular correlation in the ⁶He β decay (see Fig. 2): a transparent trap using wires to form hyperboloidal surfaces [5], a trap made of 6 rings [6] and a simple cylindrical trap presented and compared to the other geometries in [7].

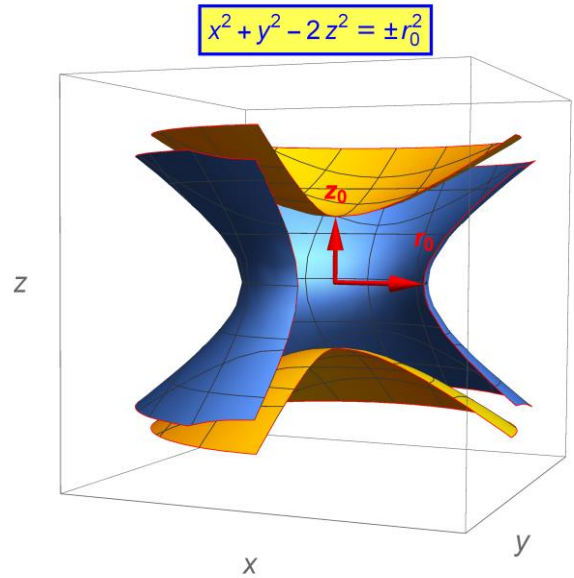

**Figure 1 : Hyperboloids commonly used for the geometry of electrodes in 3D Paul traps. The distances of the ring and end cap electrodes to the center of the trap are respectively called $r_0$ and $z_0$. They are related by the formula: $z_0 = \frac{r_0}{\sqrt{2}}$.**

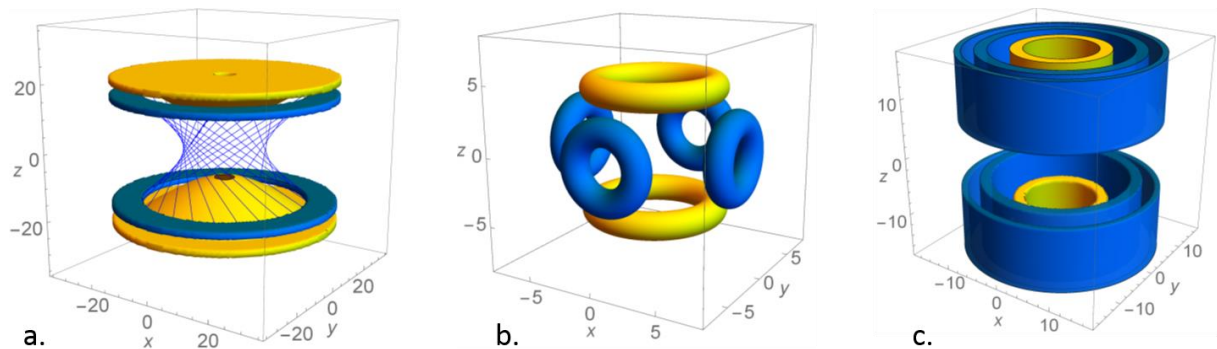

**Figure 2: Geometries originally considered for the open trap of LPCTrap. Dimensions are in mm. The electrodes mimicking the ring of Fig. 1 are shown in blue, the ones mimicking the end caps are shown in yellow. Inset a.: Paul trap whose ring was made of wires [5]. This design was inspired from a setup used at Ayme Cotton Laboratory by J. Pinard. Inset b.: trap made of rings studied in [6], inspired from a design found in [8]. Inset c.: the so-called "tube trap" [7], which was developed and finally adopted for LPCTrap.**

The so-called "tube trap" was preferred because of its higher transparency and simplicity. In such a trap, the axial symmetry and the planar symmetry reduce the multipole expansion of the potential of

Eq. (3) to a sum of zero order ($m = 0$) and even degree ($2n$) associated Legendre polynomials, respectively:

(4) $\quad V(\rho, \theta, t) = V_{rf}(t) \cdot \sum_{n=0}^{\infty} A_{2n} \left(\frac{\rho}{r_0}\right)^{2n} P_{2n}(\cos\theta)$

where the coefficient $A_{2n}$ are given by $A_{2n} = \text{Re}(C_{2n0})$, and the $P_{2n}(\cos\theta) = P_{2n}^0(\cos\theta)$ are Legendre polynomials of degree $2n$. In cylindrical coordinates, Eq. (4) can be rewritten as a sum of polynomials of even degrees in $r$ and $z$:

(5) $\quad V(r, z, t) = V_{rf}(t) \cdot \sum_{n=0}^{\infty} H_{2n}(r, z) = V_{rf}(t) \cdot \sum_{n=0}^{\infty} \frac{A_{2n}}{r_0^{2n}} \sum_{k=0}^{n} a_{2k}^{2(n-k)} r^{2k} z^{2(n-k)}$

where the $a_i^j$ coefficients are defined by the following recurrence relation to satisfy the Laplace equation:

(6) $\quad a_{i-2}^{j+2} = -\frac{i^2}{(j+1)(j+2)} a_i^j$ with $a_{2n}^0 = P_{2n}(0)$

Using Eq. (5) and (6), we observe that $V_{ideal}(r, z, t) = V_{rf}(t) \cdot H_2(r, z)$. The original trap geometry, fully axisymmetric, is shown in detail on the left inset of Fig. 3. In this geometry, the central tubes are mimicking the end cap electrodes of Fig. 1. It was optimized irrespective of the trap environment. The equipotential lines are calculated by SIMION [9] for 1 V set on the electrodes. The size and position of the tubes were finely tuned to favor the formation of a quadrupolar potential, thus minimizing the multipoles of higher order: $H_2 \gg H_{2n} \ \forall n > 1$ in the region of interest. The inset i of Fig. 4 is a zoom on the potential created in the central region of the trap, delimited by the large dashed rectangle (B) in Fig. 3. The reference potential is taken at the center of the trap. A chi-square adjustment of potential of the form of Eq. (5) yielded the harmonic coefficients as shown in Tab. 1.

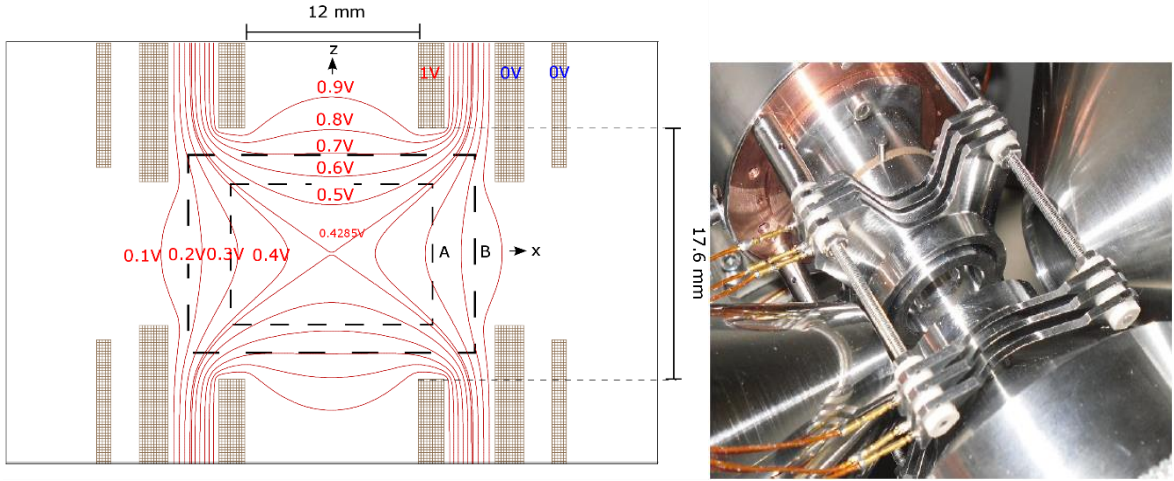

**Figure 3** : Geometry of the open Paul trap. Left inset: cross section of the trap and equipotential lines as calculated by SIMION. The z-axis is the axis of revolution of the trap. The distance between the electrodes mimicking the end caps is 17.6 mm, so that $z_0 \approx 8.8$ mm. Right inset: photograph of the trap and its environment.

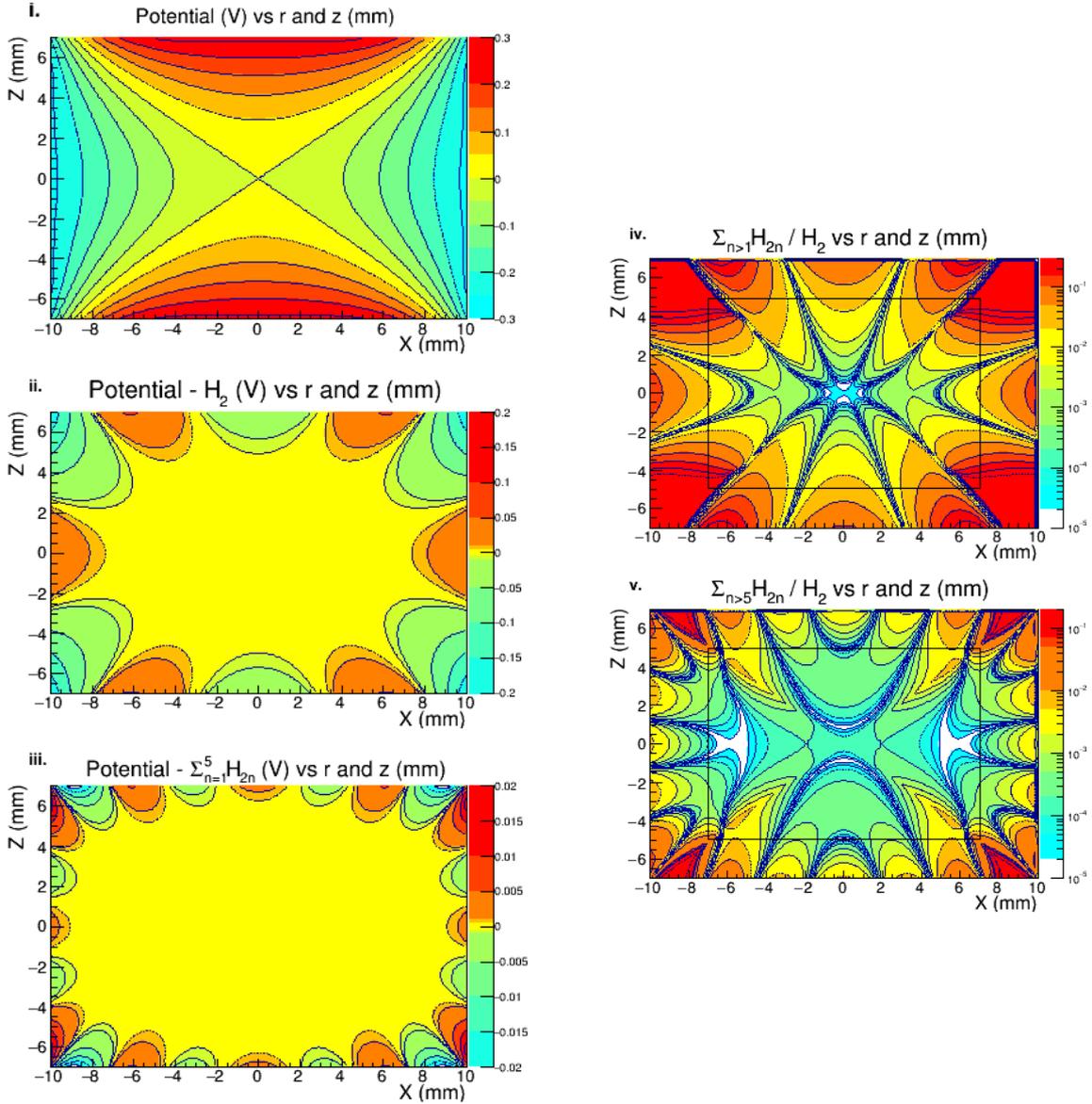

**Figure 4 :** Inset i: potential created by the electrodes in the configuration of Fig. 3. Insets ii and iii: residuals resulting from the subtraction of the different sum of multipoles to the potential of inset i. Insets iv and v: corresponding relative residuals, normalized to $H_2$. The black rectangles show the region discussed in the text, corresponding to the small dashed rectangle (A) in Fig. 3.

| Multipole | Quadrupole | Octupole | Dodecapole | 16 - pole | 20 - pole |
|---|---|---|---|---|---|
| $r_0$ (mm) | $A_2$(V) | $A_4$(V) | $A_6$(V) | $A_8$(V) | $A_{10}$(V) |
| 12.875±0.069 | 1 | -0.102±0.003 | -0.90±0.02 | -0.0661±3 | 0.51±0.01 |

**Table 1 : Coefficients of the main harmonics of the open trap potential.**

The distance parameter $r_0$ in Table 1 is consistent with a rough estimate from the distance of the central tube tips to the center of the trap (Fig. 3): $r_0 = z_0 \cdot \sqrt{2} \approx 8.8 \times \sqrt{2}$ mm. Residual fields obtained after subtracting the quadrupolar field alone and the sum of harmonics as given in Table 1 are shown in the insets ii and iii. They are normalized to the $H_2$ multipole in insets iv and v. As can be seen in the inset iv, the contributions of the octupolar and dodecapolar harmonics are dominant in the rim of the region of interest, and decrease when approaching the trap center. Their relative contribution to the potential is below 10 % in a region delimited by $r=\pm 7$ mm, $z=\pm 5$ mm around the center of the trap (black rectangle in the inset iv). In the same region, the maximal contribution of higher order multipoles (inset v) is in the order of 1%. These potentials are not taking into account the trap environment, which consist of injection, extraction optics and collimators of the recoil ion and electron detectors visible in the right inset of Fig.3. These elements, in the immediate vicinity of the trap electrodes, will affect the axial and planar symmetries of the potential. They cannot be taken into account in the reduced multipole decomposition given in Eq. 4 and 5. With such trap, embedded in the experimental environment of the LPCTrap experiment, a trapping lifetime ranging from 100 ms to 500 ms was experimentally observed [1]. The trapping losses are attributed to ion – neutral collisions and electric field imperfections [10]. The cloud temperature was determined to be of the order of 0.1 eV and the size of the order of 2.4 mm FWHM for $^6$He [1,11]. A space charge capacity of the order of 2.5 $10^5$ ions per bunch was observed with $^{39}$K [1]. For the purpose of future experiments, it is worthwhile to investigate the origin of the trapping losses, as well as possible ways of reducing the cloud size and temperature. The latter have been found to be the dominant sources of systematic uncertainty in a comprehensive analysis of the first measurement of the β-ν angular correlation parameter in the decay of $^6$He$^+$ ions, undertaken with reduced statistics [11]. In turn, they will certainly matter in the *D*-correlation measurement. In the next section, the performances of the original trap design are investigated by means of numerical investigations using the aforementioned multipole decomposition. The impact of the environment on the trap performances is then evaluated by comparing of the results with measured properties of the ion cloud. Sec. 4 investigates possible ways for enhancing the trapping performances for the MORA project.

## 3. Numerical investigations of the trapping performances

The trapping performances were investigated numerically using the expression of the potential as formulated in Eq. (5), and a Bulirsch-Stoer integration method [12] combined with a hard sphere collision model to integrate the ion motion differential equations. As a simple estimate, we use a geometrical collision cross section of the form $\pi(r_{He} + r_{K+})^2$, with $r_{He} = 30$ pm the Van Der Waals radius of He, and $r_{K+} = 140$ pm the ionic radius of K. As will be shown later, this approximation is in practice a good estimate of a more realistic cross section, which permits to reproduce mobility data available for example in [13]. Such a simple approach is meant to give qualitative results and is compared for consistency to experimental data presented in [1,11]. The work presented here will be followed by a more detailed study using the simulation package developed for the analysis of the β-ν angular parameter [14], which is able to account for space charge effects and different realistic ion – neutral interaction potentials.

### 3.1 Test cases

As a reference study, $^{39}$K$^+$ ions for which experimental data exists at LPCTrap [1] are considered. Ions are injected in the trap center with a small spatial spread $\sigma_x=\sigma_y=\sigma_z=0.1$ mm and a quite typical energy spread from a RFQ cooler buncher $\sigma_E =1$ eV. Even tiny, such a phase - space volume overcomes the phase – space capacity of the trap, as a large fraction of the ions (>50% in the most favorable cases) is lost in the first µs. Ions are considered as lost when their trajectory crosses the border of the region shown in Fig. 3, entering areas close to the electrodes where high order harmonics ($2n\geq10$) dominate, or of a downscaled region for the study of the effective trapping area as detailed in the next section. In the following, the ideal potential refers to a pure quadrupolar potential based on the $r_0$ and $A_2$ values of Table 1, while the realistic potential additionally includes the higher order harmonics up to $2n=10$. When the hard sphere collision model was used, an arbitrary high residual pressure of $2\cdot10^{-4}$ mbar of He gas was assumed, using the geometrical cross section as defined above for estimating the probability of collision. Unless specified otherwise, an ambient gas temperature of 300 K was generally adopted. In these conditions, the collision frequency was kept rather high (of the order of 1 kHz) to limit the computation time, but well below the ion macro-motion frequency, yielding results which are scalable with the lower pressures usually used at LPCTrap, and/or with more realistic collision cross sections. In the results detailed in the following, the ion cloud phase – space reaches a steady state after a few 10 µs only when no collisions are applied, and after a few 10 ms (about 30 collisions) when buffer gas cooling is applied. The trapping of $^{39}$K$^+$ ions was considered for a variable RF voltage and a fixed frequency of 600 kHz. Doing so, the stability diagram [3] was scanned for Mathieu parameter $q_z$ ranging from 0.14 to 0.83, where

(7) $\qquad q_z = \frac{4qV_0}{mr_0^2\Omega^2}$

with $m$ and $q$ the mass and charge of the ions.

The motion in an ideal Paul trap is stable for $q_z<0.908$, condition which could be numerically verified. For each individual set of parameters, the trajectories of an ensemble of 1500 ions were calculated for trapping times up to 500 ms.

### 3.2 Results
#### 3.2.1 Effective trapping region

Away from the trap center, harmonics of order larger than 2 will progressively make the real potential depart from the ideal one, eventually yielding instabilities for the ion motion, as a result of non linear resonances in the stability diagram ejecting ions from the trap (see for example [4]). In order to estimate the effect of the departure from the ideal potential on the trapping efficiency, the loss criterion described above, *i.e.* $r > 10$ mm and $|z| > 10/\sqrt{2}$ mm, corresponding to the border of the region shown in Fig. 4, was systematically downscaled by steps of 1 mm down to $r > 2$ mm and $|z| > \sqrt{2}$ mm. This loss criterion was applied to ions either injected in the ideal or in the real potential. For these simulations the buffer gas cooling was not applied. The result of this systematic investigation is shown in Fig. 5 for a Mathieu parameter $q_z=0.277$, where $^{39}$K$^+$ ions were considered as trapped when they were still flying after an arbitrary time of 10 ms: in the absence of collisions, it was in fact observed that the ion losses essentially occur in the first 50 µs following the injection in the trap, while after a flying time of about 100 µs most of the trapped ions remain indefinitely in the trap. As can be observed in Fig. 5, the number of trapped ions for the real and ideal potentials is very similar up to a radius of about 7 mm, indicating that for ions whose initial trajectory is contained in the corresponding region of space, the motion remains indefinitely stable. Beyond this radius, the real potential fails to trap the ions. A very similar behavior was found for other Mathieu parameters, showing that the effective trapping area of the real Paul trap is limited to a region of radius $r_{eff} \approx 7$ mm and axial size $z_{eff} = \frac{r_{eff}}{\sqrt{2}} \approx 5$ mm around the center of the trap, irrespective to the value of $V_0$. The reduction of the size of the trapping area is a direct consequence of the high order harmonics, whose effect starts being non negligible when their contribution becomes greater than ~10%.

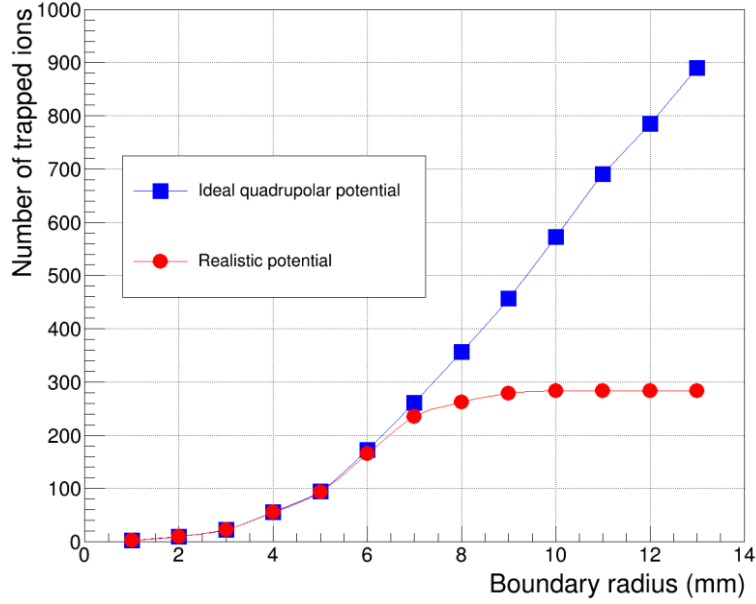

**Figure 5 : Number of trapped ions after a flying time of 10 ms versus the radius of the trap boundary for an ideal field and for a realistic field. In these simulations no collisions were considered.**

### 3.2.2 Interpretation of the buffer gas cooling and phase - space evolution

The finite boundary of the effective trapping area is expected to have some implication on the maximal energies of ions which are stored in the trap. In Paul traps, it is customary to decompose the ion motion in so-called macromotion and micromotion:

(8) $\quad z = Z + \delta$

Where $Z$ refers to the secular motion, also commonly called macromotion, and $\delta$ to the micromotion. This approximation is well established for relatively low $q_z$ values, for which a pseudo-potential well can be calculated [3]. The depth of this pseudo-potential for an ideal trap reads

(9) $\quad D_r = \frac{q_z \times V_0}{16}$

in the radial direction and

(10) $\quad D_z = \frac{q_z \times V_0}{8} = 2D_r$

in the axial direction. For a real trap, $V_{eff}$ is defined as the amplitude of the RF potential at the rim of the effective trapping region, where the quadrupolar potential is still dominating. Recalling that $r_{eff} \approx$ 7 mm is the radius delimiting this region, we get

(11) $\quad V_{eff} = V_0 \left(\frac{r_{eff}}{r_0}\right)^2$

Under these conditions one can define the associated effective pseudo-potential depths which correspond to the maximal energies of ions that can be effectively trapped [3]

(12) $\quad D_{zeff} = D_z \left(\frac{r_{eff}}{r_0}\right)^2 = 2D_{reff}$

In the present case, $\left(r_{eff}/r_0\right) \approx (7/12.9)^2 = 0.29$, so that the effective pseudo-potential depths are significantly reduced compared to an ideal trap. They approach the thermal energies of the cloud for the small values of the Mathieu parameter. In such conditions, it becomes clear that the relatively short trapping lifetimes observed with LPCTrap are caused by ion evaporation from the shallow potential of the trap. Collisions are boiling off ions from the trap when their trajectories bring them beyond the effective trapping region, or equivalently when bringing them to the high energy tail of the ion cloud. The evaporation rate will mostly depend on the ratio of the pseudo-potential depth to the thermal energies $\alpha_{Er} = \frac{D_{reff}}{kT_{eff}}$. This interpretation is corroborated by the predictions of an analytical model [15] which gives insights on the conditions of equilibration of the ion cloud. In such a model, estimates from the mean energy, cloud radius, as well as rate of evaporation are derived from the equilibrium temperature, which is found to be well approximated by

(13) $\quad T_{eff} = 2T/(1 - \frac{m_g}{m})$

in the domain of validity of the pseudo-potential approximation. In eq. (13), $T_{eff}$ is the equilibrium temperature of the ions, $T$ is the temperature of the buffer gas and $m_g$ is the mass of the buffer gas atoms. As the ratio $\alpha_{Er}$ primarily defines the evaporation rate, the thermal equilibrium is only found if $\alpha_{Er} \gg 1$. In the following, the characteristic of the ion cloud (temperature and size) and trapping lifetimes are deduced from the simulations for different buffer gas temperatures, and Mathieu parameters. The comparison of the results with the model predictions gives information on the condition of equilibration of the ion cloud. The results of the simulation are then compared to experimental results obtained with LPCTrap. Possible ways of optimizing the trap performances for the MORA project are eventually discussed in Sec. 4.

### 3.2.3 Temperature and mean radius of the ion cloud

Figure 6 presents an example of evolution of the mean ion cloud kinetic energy and radii for ions trapped and cooled by a buffer gas of He at different temperatures, and for $q_z$ =0.28 ($V_0$=66 V). Realistic potentials with harmonics $2n>2$ were used. The energies and radii are averaged over several RF cycles. In the conditions of the simulation ($P_{He} = 2 \cdot 10^{-4}$ mbar, using the aforementioned geometrical cross section), the cloud takes about 30 ms to thermalize completely. In the left inset of Fig. 6, the mean energy of ions at equilibrium is higher than the one temperature of the buffer gas, as a result from the so-called RF heating effect [16]. For comparison, the dashed lines materialize ion energies corresponding to a Brownian motion (300 K) and to the ion cloud temperature predicted by the model and given in Eq. (13): $T_{eff}$=668 K. The use of cryogenic traps at liquid He or N temperature drastically reduces the energies and radii of the trapped ions.

Figure 7 shows the evolution of the mean kinetic energies obtained by simulation for a buffer gas at room temperature for a wide range of Mathieu parameters (fixed frequency of 600 kHz but varying RF voltages). For $q_z \leq 0.2$ the energies of the trapped ions are limited by the small depth of the pseudo-potential well ($\alpha_{Er} < 1$). Under such conditions the evaporation rate is large and the lifetime of ions in the trap does not exceed a couple of collisions. For intermediate $q_z$ values ($0.2 < q_z < 0.7$) the ion cloud reaches a thermal equilibrium close to what is predicted by the model. For high $q_z$ values ($q_z \geq 0.7$) the mean energies depart from the validity of the pseudo-potential approximation. For these values, the ion motion includes other harmonics than the simple decomposition in macro- and micromotions of Eq. (8), resulting in a hotter effective temperature. For cryogenic temperatures, the evaporation rate is strongly inhibited by the large $\alpha_{Er}$ ratios. Under these conditions, the ion cloud can reach a thermal equilibrium even with a low pseudopotential well depth $D_{reff} = 0.085$ eV for $q_z = 0.14$, for which the pseudo-potential approximation is excellent, as shown in Fig. 8.

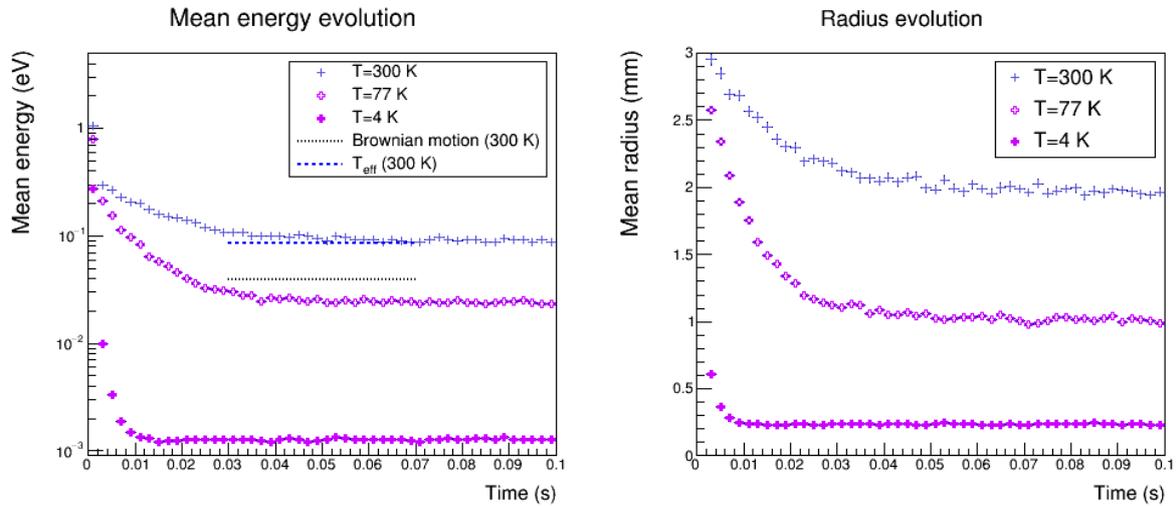

Figure 6 : Phase - space evolution of the ion cloud. Left: average ion kinetic energies for $q_z$ =0.28 and different gas temperatures. The theoretical average energy for a Brownian motion at 300 K (~39 meV) is shown by a dashed line for comparison, as well as the average energy corresponding to T=300 K from the model developed in [15]. Right: average ion radius to the center of the trap for $q_z$ =0.28 and different gas temperatures. See text for more explanations.

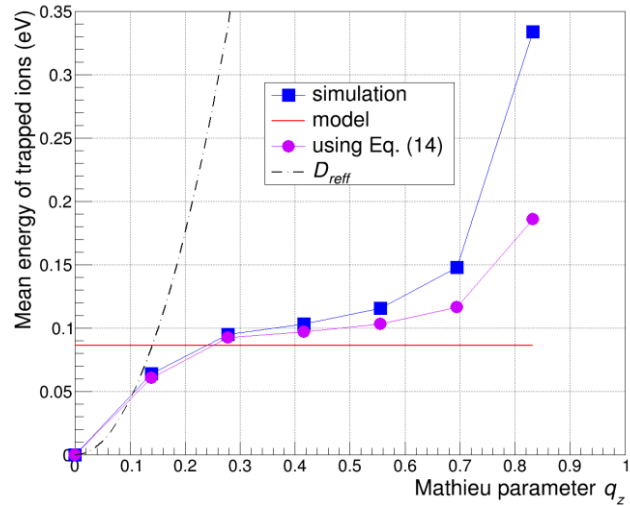

Figure 7 : Mean kinetic energy from simulations, model, and injecting the simulated average radii in Eq. (14). At low $q_z$ values the mean kinetic energy is limited by the weakness of the pseudo-potential. At high values the validity of the pseudo-potential approximation is fading out. See text for more details.

Figure 9 presents the average squared radii for different $q_z$ values. It is well known that cloud radii are related to the mean energy of the ion cloud. In [15], one finds:

$$(14) \quad \overline{E_k} \approx \frac{8eD_r}{3} \cdot \frac{\overline{\rho^2}}{r_0^2}$$

Where $\overline{E_k}$ is the average ion energies, and $\overline{\rho^2}$ the average squared radius. As a consequence, the radius decreases with higher pseudopotential well depths, so that smaller clouds could be obtained at a given temperature by using higher $q_z$ values than those presently employed for LPCTrap ($q_z \leq 0.3$), diminishing additionally the high evaporation rate.

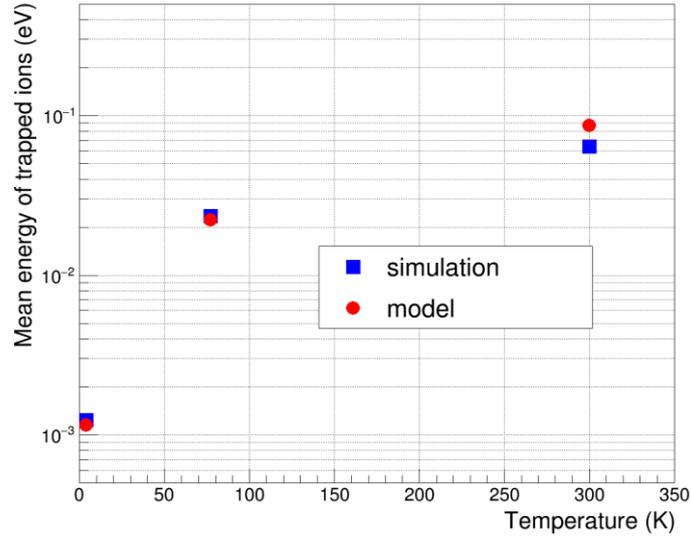

Figure 8: Comparison of the mean energies of ions from the model and simulations for a $q_z$ value of 0.14 at different temperatures. For T=300K, the shallow pseudo-potential depth does not prevent ion evaporation, thus reducing the effective temperature of the cloud.

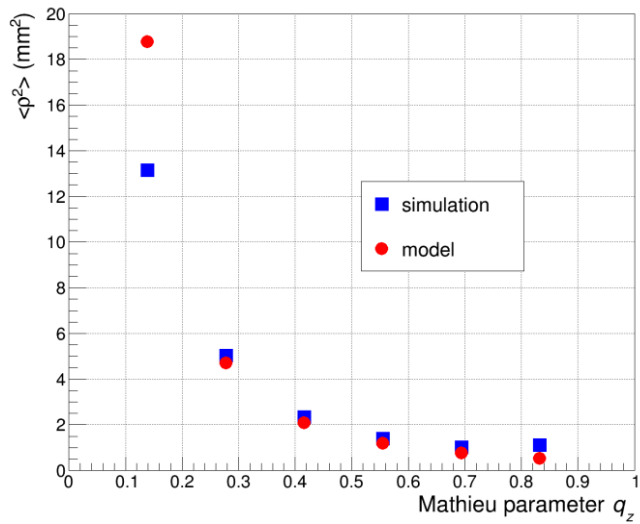

Figure 9 : Average squared radius vs $q_z$, as calculated with the Monte Carlo simulation, and predicted by the model, using Eq. (13) and (14).

### 3.2.4 Trapping lifetime

Figure 10 shows an example of lifetime plot obtained for an RF voltage of 33 V, corresponding to $q_z$=0.14. When no cooling is applied, most ions which are stable for the first 100 µs remain stable over the 500 ms of the simulation, in accordance with the observations done when determining the effective trapping region (see Sec. 3.2.1). When cooling is applied (gas at 300 K), ions are getting lost with time. These losses disappear in the case of a cryogenic Paul trap at 77 K or 4 K. The 4 K trap even enhances the trapping efficiency of ions freshly injected in the trap.

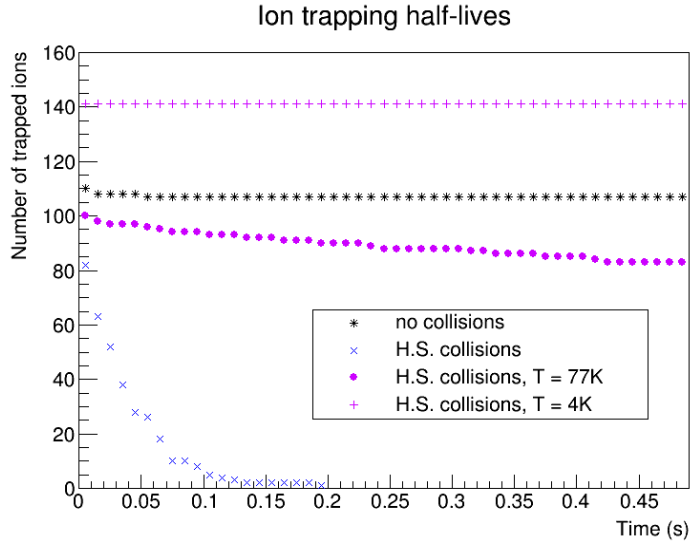

Figure 10 : Time evolution of the number of trapped ions.

The trapping average lifetimes expressed in terms of number of collisions have been determined by simulations for a number of RF voltages, as shown in Fig. 11. For the shortest half-life, the error bars correspond to fluctuations observed for simulations repeated with different starting temperatures. As it clearly appears, the half-life primarily depends on the ratio of the maximal confined ion energy to the thermal energy in the trap, $\alpha_{Er}$, defined Sec. 3.2.2. This ratio can be efficiently increased, and the evaporation rate drastically reduced, either by increasing the RF voltage (and therefore $D_{reff}$) or by reducing the temperature of the trapped ions by using gases at cryogenic temperatures. The analytical model gives best predictions for an effective radius of $r_{eff} = 7$ mm. This radius is consistent with the dimensions found for the effective trapping area, and corresponds to the limit at which the curve connecting the blue squares of Fig. 5 departs from the one connecting the red disks. For too high $q_z$ values ($q_z \geq 0.3$) the average lifetime is so large that it becomes difficult to evaluate by simulations.

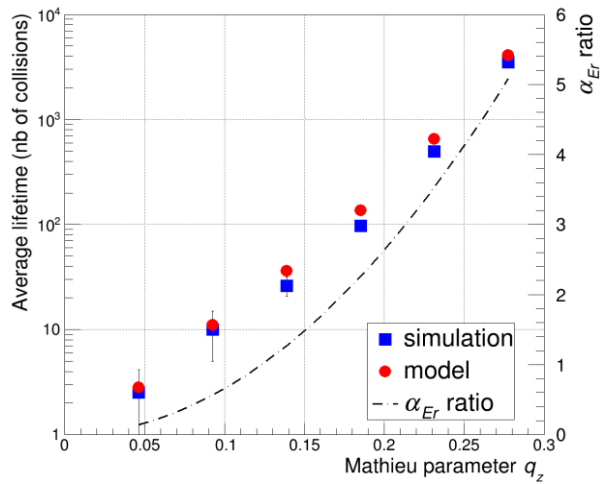

Figure 11 : Trapping half-lifes, expressed in number of collisions, as obtained from Monte Carlo simulations and from the model developed in Ref. [15]. The ratio $\alpha_{Er} = \frac{D_{reff}}{kT_{eff}}$ on which the lifetime primarily depends is also plotted as the dash line.

### 3.2.5 Comparison with experimental results

The results of the simulations discussed in Sec. 3.2.1 to 3.2.4 are based on the multipole decomposition of the potential generated by the original trap geometry, presented in Fig. 3. They do not account for the trap environment. In the immediate vicinity of the trap, injection and extraction optics, and detector collimators, visible in the right inset of Fig. 3 and shown in Fig. 1 of Ref. [17], affect the trapping potential. These elements induce potential asymmetries, which cannot be accounted for in the reduced multipole decomposition used in the present study (Eq. 4-5). We therefore use the comparison of the results of the simulation of the trap alone, as done in the previous sections, with experimental measurements of the ion cloud properties, in order to evaluate the effects of the trap environment on its performances.

Experimental results with buffer gas cooling of $^6$He and $^{39}$K are reported for example in Refs. [1] and [11]. In Ref. [11], the radial dimensions and the energy distribution of the $^6$He cloud are inferred from the measurement of recoil ion time-of-flight of $^6$Li$^+$ ions extracted from the trap and its comparison to SIMION [9] simulations including the surrounding elements of the trap and using realistic collision potentials, as detailed in Ref. [17]. Fig. 12 shows the results of the simulation for the same system.

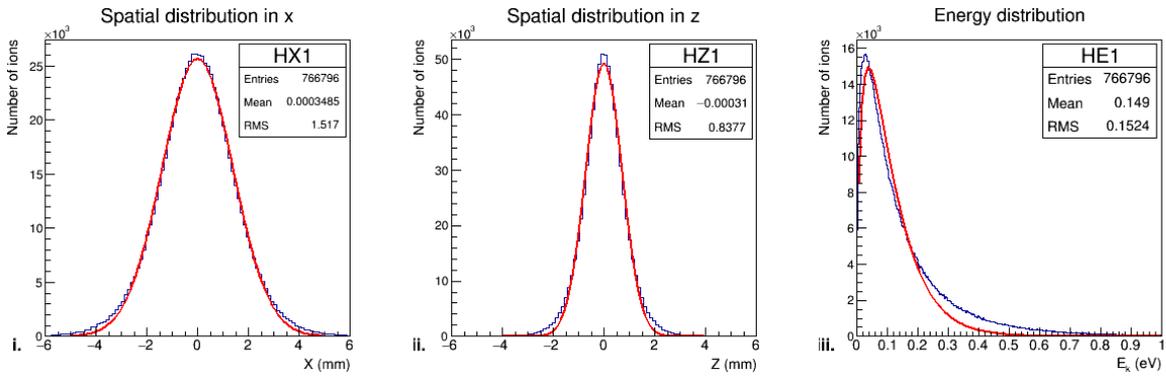

Figure 12 : From left to right: distributions along (i) the radial x dimension (ii) the axial dimension and (iii) as a function of the kinetic energy for $^6$He trapped with $V_{rf}$=60V, $\Omega_{rf}$=1.15MHz. Results of a Gaussian fit and of a Maxwell-Boltzmann distribution in red are superimposed for a comparison. The distribution in y is not shown as it is very similar to (i).

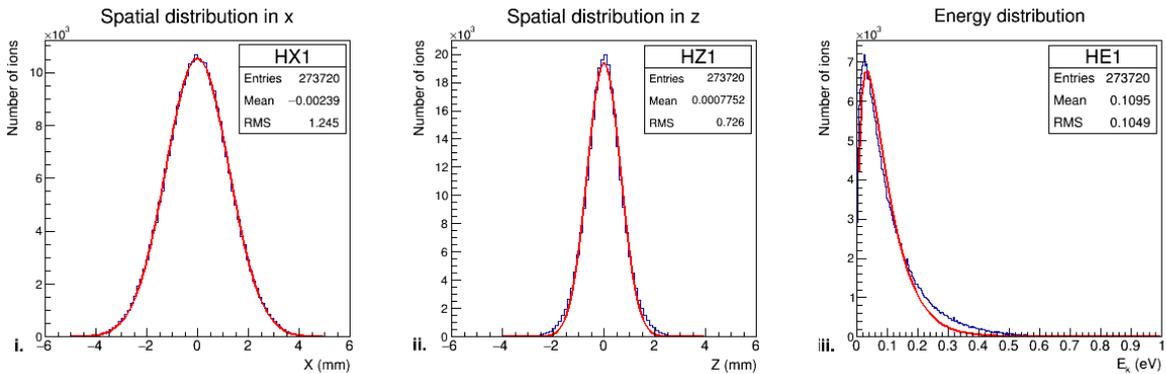

Figure 13 : Same as Fig. 12 but for a trapping radius limited to 4.5 mm.

Spatial distributions are gaussian-like, with a rms deviation in z, which is twice lower than in the radial dimensions, as the confinement forces are twice higher in this direction. The energy distribution is nearly Maxwellian. Tab. 2 enables a rapid comparison of the simulations results, model predictions and measurements. Owing to the rather high $\alpha_{Er}$ ratio ($\alpha_{Er} = 4.2$), the evaporation of $^6$He$^+$ ions marginally affects the temperature and dimensions of the cloud. With a Mathieu parameter $q_z$ = 0.44, the ion motion starts including higher order harmonics than those considered in the pseudo-potential approximation. Under these conditions, the resulting average energy and rms radius of the ion cloud are, as expected, larger than the model predictions. The simulated values are also 50% larger than the experimental values. In [14], it is shown that the hard sphere approximation yields equilibrium states (average energy and dimensions) in good agreement with more realistic atom – ion interaction potentials. The comparison of the simulated properties with the measurements therefore clearly pinpoints the role of elements surrounding the trap in creating unwanted potential harmonics, reducing further the effective trapping radius. Following a similar procedure as shown in Sec. 3.2.1, we find by downscaling the trapping radius by steps of 0.5 mm that a trapping radius of ~4.5 mm permits to match the experimental observations. The ion cloud size and mean energy for such a trapping radius are shown in Fig. 13 and listed in Tab. 2. This is a significant reduction compared to the effective radius of 7 mm determined in Sec. 3.2.1 for the original trap geometry. Consequently, a particular attention will have to be paid to the design of the environment of the trap for the MORA project, in order to preserve the trapping potential from induced asymmetries, and unwanted multipoles.

|  | Mean square radius $r^2$ (mm²) | Mean energy (eV) |
|---|---|---|
| Simulations | 2.30 | 0.149 |
| Model | 1.90 | 0.116 |
| Experimental values [11,17] | 1.5 | 0.107(7) |
| Simulations (4.5 mm radius) | 1.55 | 0.110 |

Table 2: Comparison of mean square radius and energies obtained at equilibrium with the simulation presented here, with the model predictions and the experimental measurements [11,17]. The experimental values can be reproduced by downscaling the effective trapping radius of the original trap geometry to ~4.5mm.

As the hard sphere approximation yields correct equilibrium states [14], it is also expected to give a correct qualitative description of the ion cooling and evaporation processes. In order to probe the accuracy of the description of these processes, the cooling times obtained in Sec. 3.2.3 are compared to results of a Monte Carlo simulation using a realistic potential derived from a quantum microscopic calculation [18]. For the latter, we consider an interaction potential for the HeK[+] system of the form:

$$(1) \quad V_{int}(r) = A \cdot (C_n r^{-n} + C_6 r^{-6} + C_4 r^{-4})$$

The potential with the best sets of parameters obtained for n=11 is shown in Fig. 14. For the average energies of the trapped ions, the predominant term will clearly be $C_n r^{-n}$. The distance of closest approach yields a comparable estimate for the hard sphere radius than the one introduced above, about 4 Bohr radii. Fig. 14 compares drift velocities and mobility data to what could numerically be obtained using the realistic potential reduced to its repulsive core term $C_n r^{-n}$. In Fig. 15, E/N is the ratio of the electric field, E, to the neutral gas concentration N, using the same conventions as in [13]. As expected, for low values of E/N, the model overestimates the experimental data, but at high values of E/N, which will matter for LPCTrap (E/N~$10^4$), the results are in correct agreement. Finally, Fig. 16 shows the thermalisation time inside the Paul trap for a value of the Mathieu parameter $q_z$ = 0.28 and for a gas pressure $P_{He} = 10^{-4}$ mbar. We find a thermalisation time of 70-80 ms, about two times longer than what was obtained with $P_{He} = 2 \cdot 10^{-4}$ with the hard sphere model (Fig. 6). Here again, the hard sphere approximation agrees reasonably well with the description of collisions using realistic interaction potentials, even when using a rather crude cross section estimate.

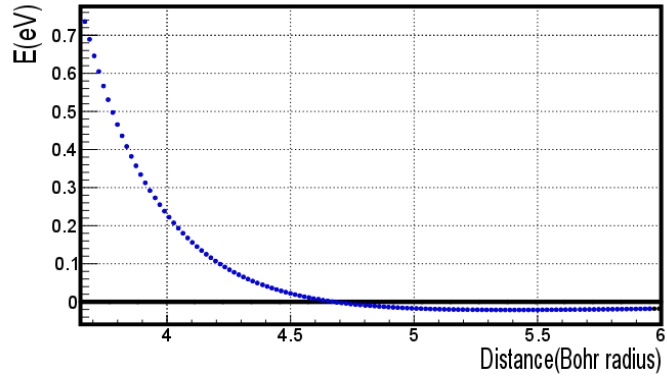

Figure 14 : Realistic interaction potential derived from [18] for the HeK$^+$ system.

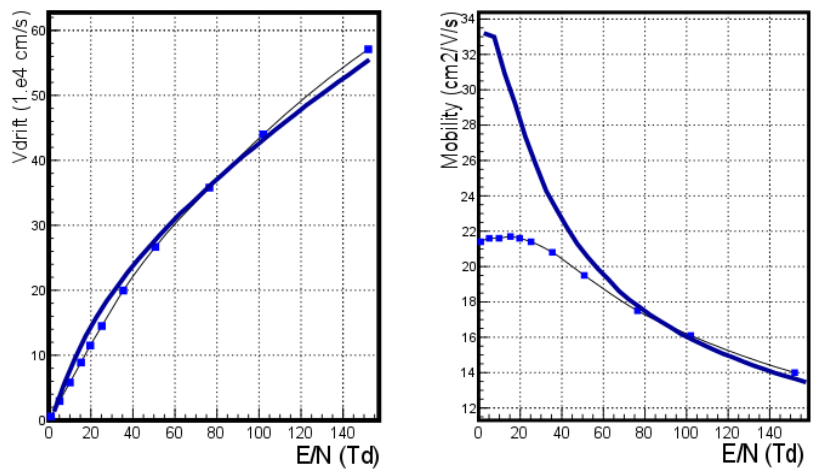

Figure 15 : Mobility data (blue dots) compared to the predictions of a Monte Carlo simulation using the realistic potential for the HeK$^+$ system detailed in the text (bold line). The left inset shows the drift velocities while the right inset shows the deduced mobilities. E/N is the ratio of the electric field, E, to the neutral gas concentration N, using the same conventions as in [13].

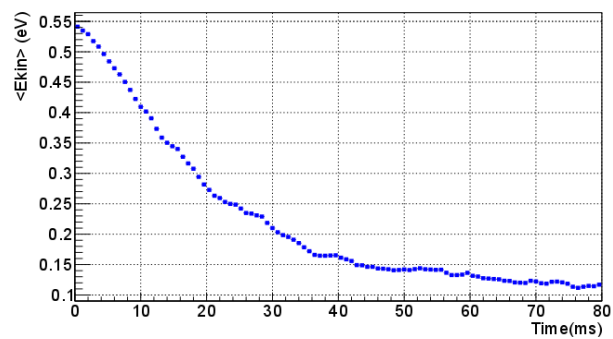

Figure 16 : Energy of the ions as a function of time, as calculated using the realistic interaction potential shown in Fig. 13, for a Mathieu parameter of $q_z$ = 0.28 and a gas pressure $P_{He}$= $10^{-4}$ mbar. The ion motion is fully thermalized after 70-80 ms.

In Ref. [1], typical thermalisation times of 20 ms, and trapping lifetimes of 500 ms are reported for all ion species, for a He pressure of 1.5·$10^{-5}$ mbar. As an alkali ion, $^{39}$K$^+$ does not likely recombine with surrounding atoms or molecules, so that the observed trapping lifetime will essentially be due to the evaporation of ions. If the pressure gauge reading gives a correct estimate of the buffer gas pressure in the trap, both characteristic times are about one order of magnitude lower than expected from simulations, using a downscaled trapping radius of 4.5 mm for estimating the evaporation rate. A consistent observation is done with the cooling time of $^6$He$^+$ ions in H$_2$ reported in [11]: only a pressure one order of magnitude higher than the quoted $P_{H2}$=2·10$^{-6}$ mbar would be able to account for the observed thermalisation time of 20 ms. Table 3 summarizes the experimental observations and estimated values from simulations. As we have built some confidence in the accuracy of the hard sphere approximation presented here, which was satisfactorily compared to simulations using realistic interaction potentials, these observations pinpoint a probable bias in the experimental pressure estimate.

| System | Measured buffer gas pressure | Experimental cooling time | Estimated cooling time from simulations | Experimental trapping lifetime | Estimated trapping lifetime from simulations |
|---|---|---|---|---|---|
| $^6$He$^+$ + H$_2$ | 2·10$^{-6}$ mbar | 25 ms | 400 ms ~10 collisions | >0.1s | 4 s ~100 collisions |
| $^{39}$K$^+$ + He | 1.5·10$^{-5}$ mbar | 20 ms | 350 ms ~16 collisions | 0.5 s | 5 s ~230 collisions |

Table 3: Summary of cooling and trapping lifetimes as experimentally observed, and deduced from simulations. The simulations assume a trapping radius limited to 4.5mm, as discussed above, to evaluate the rate of evaporation. The cooling and trapping lifetimes are defined as the times after which the ion cloud is either fully cooled or fully evaporated.

With the results presented in Sec. 3, some conclusions can be drawn for the realization of future experiments, where a fine control of the properties of the ion cloud (size and energies) and best performances in term of capture and trapping times are desired. This is particularly the case for the MORA project.

## 4. Prospects for the MORA project

### 4.1 The MORA project

LPCTrap provided new data on the β-ν angular correlation in the decay of $^6$He, $^{35}$Ar and $^{19}$Ne [1,19]. These experiments are currently being analyzed. The $^6$He data should permit to set competitive constraints on the existence of tensor currents. The $^{35}$Ar and $^{19}$Ne data should permit to improve the determination of the first element of the CKM matrix, $V_{ud}$, from mirror nuclides. As a rather natural extension of the physics program of LPCTrap, the measurement of the triple correlation *D* is one of the most promising experiments to search for New Physics. The *D* correlation violates Time reversal symmetry, and via the CPT theorem, is sensitive to CP violation. Such a measurement, undertaken in the frame of the *Matter's Origin from the RadioActivity* (MORA) project [20] makes again use of the trap transparency to detect in coincidence the recoil ion and beta emitted during the decay. MORA will use an innovative in trap polarization method, permitting a very efficient orientation of large samples of $^{23}$Mg and $^{39}$Ca ions. Isotopes will be laser polarized as ions in the trap, similarly to what is done at β-NMR setups like COLLAPS (see eg [21, 22]). We will first focus on $^{23}$Mg, which is relatively easy to laser polarize. Recent studies showed that pulsed lasers have to be preferred to CW ones because of their larger bandwidth, which covers the Doppler broadening of the transitions of interest (typically 5 GHz as compared to 200 MHz respectively) due to the thermal agitation of the ion cloud. The proof-of-principle of the polarization and of the *D* correlation measurement with $^{23}$Mg$^+$ ions will be undertaken at JYFL, before moving the experiment setup back to GANIL [20].

### 4.2 Important parameters for MORA

With a mean ion kinetic energy of 0.1 eV in the trap, the pulsed laser bandwidth matches well the Doppler broadening of the trap. In the trap, it is expected that the achievable polarization degree will be close to 100%, after a time, which will depend on the laser power overlapping with the ion cloud. The rapidity of the polarization will therefore depend on the radius of the ion cloud. With a radius of 1-2 mm, the estimated beam power will be enough to reach a full polarization after a few ms [20]. Such parameters are well within the typical performances of LPCTrap (Sec. 3). Considering the 11.3 s half-live of $^{23}$Mg, the critical parameter for the MORA experiment is the trapping lifetime, presently limited to 100-500 ms. Such a short lifetime would be a source of unwanted background and/or of unwanted losses for the *D*-correlation measurement. A more efficient cooling mechanism than presently used at LPCTrap is therefore envisaged for MORA to reduce the trapping losses.

### 4.3 Possible optimizations

Using the results of the numerical study and of the model detailed in Sec. 3, the experimental trapping lifetime which was found to be of the order of 500 ms, could be extended to a few seconds in two ways (see Fig. 11 and discussion of the importance of the $\alpha_{Er}$ factor):

- Increasing the depth of the effective pseudo-potential well by
  - using a higher RF voltage, and/or
  - extending the effective trapping area
- Cooling the ion cloud down to cryogenic temperatures

Both options would additionally permit to trap more efficiently freshly injected ions (Sec. 3). Nevertheless, increasing the RF voltage alone to increase the depth of the effective pseudo-potential well would have an unwanted side effect: the maximal recoiling energy of the $^{23}$Mg ions being of the order of 300 eV only, a higher RF field would disturb more the recoil ion trajectories than a lower field. This would result in comparably higher systematic effects [11] to be cared for in the analysis of the experimental data and a possible washout/loss of sensitivity of the *D*-correlation measurement. It is therefore much more advantageous to extend the effective trapping area, by fine tuning the shape of the transparent Paul trap in order to suppress even further the contribution of harmonics of degrees larger than 2. This optimization is actively undertaken by the MORA collaboration using methods more sophisticated than presented in Sec. 2, which will be detailed in future publications.

Cryogenic Paul traps have already been built for different purposes (see for example [23,24]) and could eventually be considered for the *D*-correlation measurement, in case the present optimizations would still yield too short trapping lifetimes. Alternatively, the use of sympathetic cooling could be envisaged. The laser cooling of Ca$^+$ ions in a trap very similar to LPCTrap was recently demonstrated for the TRAPSENSOR project [25,26]. $^{23}$Mg$^+$ or $^{39}$Ca$^+$ ions could then be cooled by collisions with laser cooled Ca$^+$ ions. This cooling scheme will be studied by the TRAPSENSOR team.

## 5. Conclusion

The open LPC Paul trap uses a simple geometry which has been optimized in order to minimize the contribution of non quadrupolar harmonics in an extended region of the electrodes inner space. The design has been successful for performing the β−ν angular correlation measurements at GANIL, and is now further used for the commissioning of the TRAPSENSOR project. An optimization of the trap is being undertaken for a measurement of the *D* correlation in the beta decay of oriented nuclei in the frame of the MORA project. Simulations have been carried out to understand the present limitations. The numerical studies have shown that the trap geometry generate potential harmonics of order larger than 2 limiting the trapping region dimensions to an effective radius of 7 mm. Comparing measured properties of the ion cloud with the simulations, it is shown that the trap environment further reduces this radius to ~4.5 mm. Following this finding, the limited trapping lifetimes observed in LPCTrap are explained by the evaporation of ions from the trap because of collisions with the residual gas bringing them to the boundary of the trapping region. For the MORA project, the optimization of the trap which is presently being undertaken consists in enlarging the effective trapping area by fine tuning the shape of the electrodes, and paying particular attention to the design of the elements directly surrounding the trap. The resulting enlarged depth of the effective pseudo-potential should permit to reduce efficiently the trapping losses. In principle, the use of buffer gas cooling at cryogenic temperature would additionally permit achieving indefinitely long trapping times, while reducing efficiently the width of the ion cloud energy distribution. Another cooling method which is being investigated by the TRAPSENSOR team is the sympathetic cooling with laser cooled $Ca^+$ ions. Both cooling techniques could equally be considered for the MORA project, in case the optimization of the trap geometry would not be sufficient to enlarge the trapping lifetime to more than a fraction of the radioactive half-life of the isotopes.

## 6. Acknowledgements

The authors acknowledge the support from Region Normandie for the MORA project.